\documentclass[12pt,preprint]{aastex}
\usepackage{graphicx}

\begin{document}

\title{The Redshift Evolution of the Tully-Fisher Relation as 
a Test of Modified Gravity}

\author{Christopher M. Limbach, Dimitrios Psaltis, and Feryal \"{O}zel}

\affil{University of Arizona, Steward Observatory, 933 N.\ Cherry Ave., 
Tucson, AZ 85721}

\begin{abstract}

The redshift evolution of the Tully-Fisher Relation probes
gravitational dynamics that must be consistent with any modified
gravity theory seeking to explain the galactic rotation curves
without the need for dark matter. Within the context of
non-relativistic Modified Newtonian Dynamics (MOND), the
characteristic acceleration scale of the theory appears to be related
to the current value of either the Hubble constant, i.e.,
$\alpha\simeq cH_0$, or the dark energy density, i.e., $\alpha\simeq
(8\pi G\rho_\lambda/3)^{1/2}$.  If these relations are the
manifestation of a fundamental coupling of $a_0$ to either of the two
cosmological parameters, the cosmological evolution would then dictate
a particular dependence of the MOND acceleration scale with redshift
that can be tested with Tully-Fisher relations of high-redshift
galaxies.  We compare this prediction to two sets of Tully-Fisher data
with redshifts up to $z=1.2$. We find that both couplings are excluded
within the formal uncertainties. However, when we take into account
the potential systematic uncertainties in the data, we find that they
marginally favor the coupling of the MOND acceleration scale to the
density of dark energy.

\end{abstract}

\section{Introduction}

The relation between luminosity and spectral line widths in spiral
galaxies has been widely established as a probe of the dark matter
content of these galaxies (Tully \& Fisher 1977). This, so called
Tully-Fisher Relation (TFR), has subsequently been constructed in
various forms with the stellar mass or baryonic mass of a galaxy
serving as a proxy for luminosity, and similarly, with the maximum
rotational velocity or asymptotic rotation velocity as a substitute
for line width. Of the many parameter combinations, the tightest TFR
fits occur between baryonic mass and asymptotic rotation velocity
(Verheijen 2001).

Galactic velocity profiles that are asymptotically flat appear
ubiquitous, yet they were not expected from Newtonian gravitational
models of the baryonic matter distribution. Additional sources of the
gravitational field, such as dark matter halos (Rubin et al 1985), or
modifications to the gravitational theory (e.g., Milgrom 1984) are
necessary for their explanation. A successful explanation of the
galactic rotation curves must also be compatible with the Tully-Fisher
Relation because of the latter's dependence on the asymptotic
rotational velocities of galaxies. 

The Modified Newtonian Dynamics (MOND) paradigm was first developed as
a modification to gravity at low accelerations for spiral galaxies and
an alternative to dark matter (Milgrom 1984; for a thorough treatment
of MOND see Milgrom 2001, 2002b, 2008 and Sanders \& McGaugh
2002). One of its early successes has been the fitting of galaxy
rotation curves with a parametric function that is determined only by
the amount of visible matter in the galaxy and a single parameter
$\alpha_0$ which represents the characteristic acceleration below
which gravity becomes non-Newtonian (Bekenstein \& Milgrom
1984). 

Non-relativistic MOND modifies Newtonian gravity such that the net
acceleration experienced by a test particle in a gravitational field
is $a_{\rm MOND}=(a_{{\rm}N}\alpha_0)^{1/2}$, where $a_{{\rm}N}$ is
the Newtonian gravitational acceleration. This phenomenological model
also leads naturally to the Tully-Fisher Relation, since the velocity
$V_{\rm f}$ of an asymptotically flat rotation curve at large
distances is related to the baryonic mass $M$ of the galaxy by
(Milgrom 1984)
\begin{equation}
M=\frac{V_{\rm f}^4}{G \alpha_0}\;.
\label{TFMond}
\end{equation}

The most unexpected result of fitting rotation curves with MOND is
that the acceleration parameter is nearly the same for all
galaxies. Moreover, its value is comparable to two combinations of
cosmological parameters that have dimensions of acceleration (Milgrom
2001), i.e.,
\begin{equation}
\alpha_0\simeq c H_0 \simeq c\sqrt{\frac{8\pi G \rho_\lambda}{3}}\;,
\label{coincidence}
\end{equation}
where $H_0$ is the value of the Hubble constant and $\rho_\lambda$ is
the cosmic density of dark energy. In the context of the Cold Dark
Matter (CDM) paradigm, this similarity is believed to arise from the
cosmological evolution of dark matter halos, which naturally depends
on the cosmological expansion. Analytic attempts to account for the
MOND acceleration scale within the CDM framework have been indeed made
(Kaplinghat \& Turner 2002) but also refuted (Milgrom 2002a). 

MOND predictions have been examined in several other settings, from
clusters of galaxies (e.g., Clowe et al.\ 2006) to larger cosmological
scales (e.g., Scott et al.\ 2001). These studies have established the
difficulties of using the MOND phenomenology to account for this wide
set of cosmological phenomena. Even without focusing on these issues,
if the MOND phenomenology describes a deviation from Newtonian gravity
at galactic scales, the numerical similarity of the various physical
quantities shown in relation (\ref{coincidence}) will need to be
explained.

MOND is based on a non-relativistic phenomenological function and,
therefore, it cannot lead to predictions for the cosmological
evolution of the Hubble parameter or of the acceleration scale
$\alpha_0$. Different attempts to devise a relativistic equivalent to
MOND lead to varying predictions. For example, in TeVeS, which is a
Tensor-Vector-Scalar relativistic theory that accounts for a large
part of the MOND phenomenology, the similarity described by
equations~(\ref{coincidence}) is coincidental (Bekenstein \& Sagi
2008). On the other hand, in a recently proposed theory which attempts
to account simultaneously for the dark matter and the dark energy
problem (Zhao \& Li 2008), the acceleration scale couples to the
density of dark energy.

If the connection between the acceleration $\alpha_0$ and either the
Hubble parameter $cH$ or the density of dark energy
$\sqrt{\rho_\lambda}$ can be proven experimentally, then this will
provide important clues towards developing a successful relativistic
theory that is consistent with the MOND phenomenology. Indeed, within
general relativity, the Einstein equivalence principle guarantees that
the cosmological expansion will affect the motion of a test particle
only to second order in $(cH_0)^2$ (Peebles 1993). Therefore, the
acceleration scale $\alpha_0$ will couple linearly to the Hubble
parameter only if the Einstein equivalence principle is violated at
cosmological scales. In this case, MOND will be a modification of
inertia rather than of the gravitational field (see Milgrom 2006). On
the other hand, the acceleration scale $\alpha_0$ will couple to the
dark energy only if the latter is not a manifestation of a
cosmological constant but rather of a dynamical field that couples
non-linearly to the metric. This is true because a cosmological
constant or a minimally coupled field that describes dark energy
introduces a characteristic scale in curvature and not in
acceleration. As a result, the relativistic version of MOND in this
case will require a modification of the Einstein field equations.

The evolution of the MOND acceleration scale with redshift provides
the means with which we can test, in a phenomenological way, the
connection represented in equation~(\ref{coincidence}) (see Milgrom
2008 and Bekenstein \& Sagi 2008).  This is true because the rate of
the cosmological expansion $H\equiv \dot{a}/a$ and the density of dark
energy have distinct dependences on redshift. In this paper, we use
observations of the so-called Tully-Fisher intercept back to a
redshift of $z=1.2$ (Weiner et al. 2006; Kassin et al. 2007) as a
proxy for the MOND acceleration scale (see Eq.~[\ref{TFMond}]). This
allows us to infer the evolution of $\alpha_0$ using only the exact
phenomena that non-relativistic MOND was designed to explain, without
the need for extrapolating the phenomenology to include a treatment of
photon trajectories. We then compare the observations to the expected
evolution of the MOND acceleration scale when coupled to the Hubble
parameter or the density of dark energy, for values of the
cosmological parameters that are consistent with recent measurements
(Spergel et al. 2007).

\section{The evolution of the MOND acceleration scale}

For our calculations, we postulate that equation (\ref{coincidence})
represents a physical relationship between the MOND acceleration scale
and either the density of dark energy or the Hubble parameter. Writing
these explicitly in terms of cosmological parameters, the independent
couplings take the form
\begin{equation}
\alpha_0 = cH = c \frac{\dot{a}}{a}
\label{hubble_coupling}
\end{equation}
or
\begin{equation}
\alpha_0 = \sqrt{\frac{8\pi G}{3}\rho_\lambda} = \sqrt{\frac{8\pi G}{3}\rho_\lambda a^{-3(1+w)}}  ,
\label{de_coupling}
\end{equation}
where $a$ is the scale factor of the universe and $w$ is a parameter
that describes the dark energy equation of state,
$P_{\lambda}=w\rho_{\lambda}$.

We separately model the time evolution of the two couplings with the
first Friedmann equation,
\begin{equation}
\left(\frac{\dot{a}}{a}\right)^{2} = \frac{8\pi G}{3}(\rho_{\rm m}+
\rho_\lambda)\;,
\label{friedmann}
\end{equation}
where we assume a flat universe ($k=0$). 

As previously stated, non-relativistic MOND cannot lead to predictions
of the cosmological evolution of the Hubble parameter or the density
of dark energy. However, the Friedmann equations have been shown
through observations to be a successful predictor of cosmic evolution
beyond redshift $z=2$. In this context, we take the first Friedmann
Equation as an empirical fit of cosmological observations back to
redshift $z=2$, rather than a self-consistent physical theory. For all
our calculations, we use cosmological parameters from the third year WMAP
results (Spergel et al. 2007), $\Omega_\lambda = 0.716$, $\Omega_{mat}
= 0.268$, $\Omega_{rad} = 0.016$. We solve the Friedmann equations
using the forth order Runge-Kutta algorithm in order to obtain the
time evolution of $\alpha_0$ with the MOND couplings specified by
equations~(\ref{hubble_coupling}) and (\ref{de_coupling}).

We now consider the results of coupling the MOND acceleration scale to
the Hubble parameter and the dark energy density in turn. Figure~1 shows
the results of coupling the MOND acceleration scale to the Hubble
parameter for three different values of the dark energy equation of
state. The acceleration scale increases with redshift independent of
the equation of state, although less so for more negative values of
$w$. Figure~2 shows the same quantities with the MOND acceleration
scale coupled to the density of dark energy. For $w$ greater than
$-1$, the acceleration scale increases as in the case of coupling to
the Hubble parameter. In contrast, with $w<-1$ the acceleration scale
decreases with redshift. Only when $w=-1$ does $\alpha_0$ remain
constant in time, as would be expected from inspection of
equation~(\ref{de_coupling}).

Figure~3 shows the MOND acceleration scale inferred from high-redshift
TFR data (c.f.\ Eq.~\ref{TFMond}) along with the values predicted
using select Hubble and dark energy density couplings. The Hubble
parameter and dark energy density couplings are shown with $w=-1.00$
based on the constraint $w = -1.08 \pm 0.12$ imposed by the latest
WMAP data (Spergel et al. 2007). We also display the $w=-1.55$ dark
energy coupling line, despite the problems inherent to such a model,
for easy visual comparison to the data from Weiner et al. (2006),
shown as open circles, and Kassin et al. (2007), shown as filled
squares.

The measurements from Weiner et al. (2006) were taken by the Team Keck
Redshift Survey in the rest-frame J infrared band. They observed line
of sight dispersion with the DEIMOS spectra boxcar to a limiting
resolution of 56 km s$^{-1}$ in the rest frame. As in Weiner et
al. (2006), the local TFR intercept from Watanabe et al. (2001) is
plotted in Figure~3 for reference. These results are not corrected for
stellar formation rates. However, Weiner et al. develop a model to
predict the systematics in their data using an exponentially decaying
star formation rate with the characteristic formation time scaled to
galactic mass. With this model they predict a $\sim 0.05$ dex evolution of
the TFR intercept from $z=0.4$ to $z=1.2$.

Kassin et al.\ (2007) obtained their redshift data from the
All-Wavelength Extended Groth Strip International Survey
(AEGIS). Their photometric measurements were provided by the Hubble
Space Telescope Advanced Camera for Surveys (HST/ACS) and spectra
gathered from the Deep Extragalactic Evolutionary Probe 2 Survey
(DEEP2). Kassin et al. (2007) converted their luminosity data to stellar mass
using the method of Lin et al. (2006) but still predicted a systematic
offset of \~0.05 dex between $z=1$ and $z=0$. This arises from
considering an exponentially decreasing star formation rate in a
typical spiral galaxy, In Figure~3, as in Kassin et al. (2006), we
adopt the local M$_\star$TFR intercept as found in McGaugh (2005) with
the maximum disk method.

Within formal uncertainties, the data in Figure~3 are not consistent
with either coupling to the Hubble parameter or dark energy density
with $w=-1.00$. We find that the current high redshift Tully-Fisher
data can only be fit within formal uncertainties by adopting $w
\approx -1.55$, which lies well outside the constraints imposed by the
WMAP third year data.

The decreasing trend in the observed Tully-Fisher intercept with
redshift may be due to three sources of systematic errors: the unknown
evolution of the initial mass function, of the star formation rate, and
of the stellar population. The model of Weiner et al. (2006) predicts
a $-0.05$ dex decrease in the TFR intercept from $z=0.4$ to $z=1.2$
when taking into account all of these factors. However, this only
accounts for roughly half of the observed decrease. Kassin et
al. (2007) correct their data for evolution in stellar populations
only and add that a $-0.05$ dex decrease in the TFR intercept from
$z=0$ to $z=1$ would be typical based on exponentially decaying star
formation models.  Taking these systematic errors into account, the
data from Kassin et al. (2007) are still formally inconsistent with no
evolution. But, with systematic errors accounted for, coupling of the
MOND acceleration parameter to the dark energy density with a
realistic value of the dark energy equation of state may be consistent
with both datasets. On the other hand, unreasonable values of the dark
energy equation of state will still be necessary to make coupling to
the Hubble parameter consistent with the data. This would appear to
favor coupling to the dark energy density, but there may still be
additional systematic errors unaccounted for.

As with coupling to the dark energy density, N-body dark matter halo
simulations have predicted little to no evolution in the stellar mass
Tully-Fisher intercept after $z=1$ (Diemand et al. 2007). More
recently, the effects of gas physics have been incorporated into these
simulations, but still no evolution in the intercept is observed after
$z=1$ (Portinari \& Sommer-Larsen 2008). Simulations that include gas
physics do show significant mass evolution in individual galaxies
along with a corresponding change in rotation velocity. This has the
effect of moving galaxies along the Tully-Fisher Relation but not away
from it, thereby leaving the intercept unchanged. Therefore, the
predictions of Tully-Fisher intercept evolution within the $\Lambda
CDM$ paradigm and the dark energy coupled MOND paradigm with $w=-1$
are indistinguishable.

\section{Conclusion}

In this paper, we calculated the time-evolution of the MOND
acceleration parameter $\alpha_0$ when coupled to either the Hubble
parameter or to the density of dark energy. For a dark energy equation
of state consistent with recent data, we expect a positive evolution
of $\alpha_0$ with redshift if it couples to the Hubble parameter and
little to no evolution with redshift if it couples to the dark energy
density. The latest measurements of the Tully-Fisher Relation support
neither coupling scenario within formal uncertainties. Observations to
$z=1.2$ indicate a decrease in both the baryonic mass and J-band
luminosity Tully-Fisher intercepts, which are coupled to the MOND
acceleration parameter as shown in equations~(\ref{hubble_coupling})
and~(\ref{de_coupling}). However, the systematic uncertainties
introduced by the initial mass function, stellar formation and stellar
population evolution affect the data significantly. Sources of
systematic uncertainty tend to shift the Tully-Fisher intercept data
upward, so that the net trend has a slight downward evolution with
redshift. The corrected data would then be consistent with a coupling
of $\alpha_0$ to the dark energy density with an observationally
allowed equation of state with $w\simeq -1$. However, the
uncertainties in such a large correction for systematics also makes no
evolution in the TF intercept plausible.

Definitive conclusions can only be drawn after the systematic errors
are more precisely quantified. Future observations from large
telescopes both on the ground and in space of galaxies at higher
redshifts will be particularly helpful, since the difference between
the MOND acceleration scale evolving coupled to the dark energy
density or the Hubble parameter increases with redshift.

\acknowledgements

We thank Ben Weiner and Anthony Aguirre for many helpful
discussions. D.\ P.\ is supported by the NSF CAREER award NSF
0746549. F.\ \"O.\ is supported by the NSF grant AST-0708640.

\begin{figure*}
\centering
\includegraphics[angle=0.0,scale=0.6]{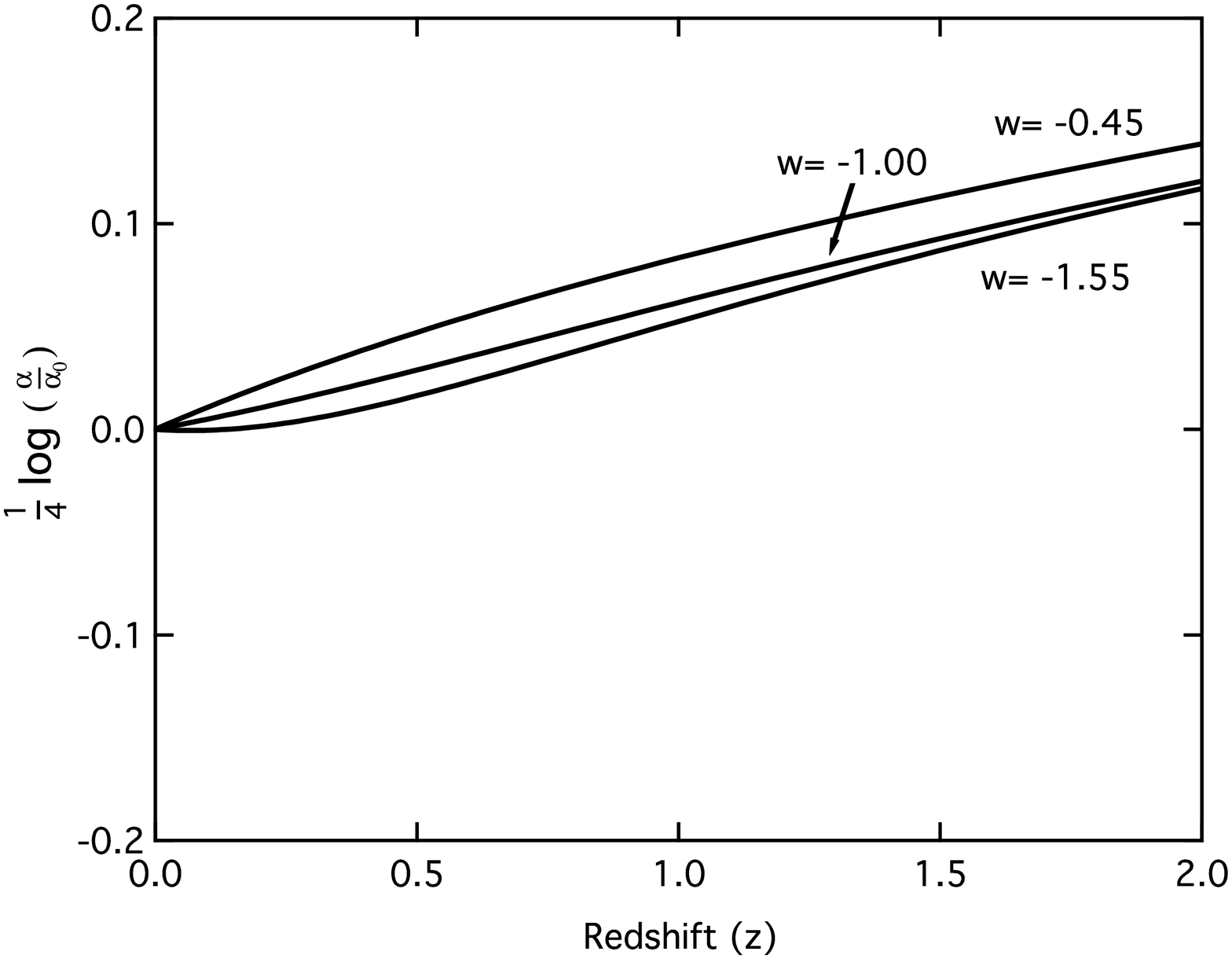}
\caption{Redshift evolution of the MOND acceleration scale normalized
to the present value $\alpha_0$ if it evolves coupled to 
the Hubble parameter for different values of the 
dark energy equation of state $w$.} 
\label{fig:hubble}
\end{figure*}

\begin{figure*}
\centering
\includegraphics[angle=0.0,scale=0.6]{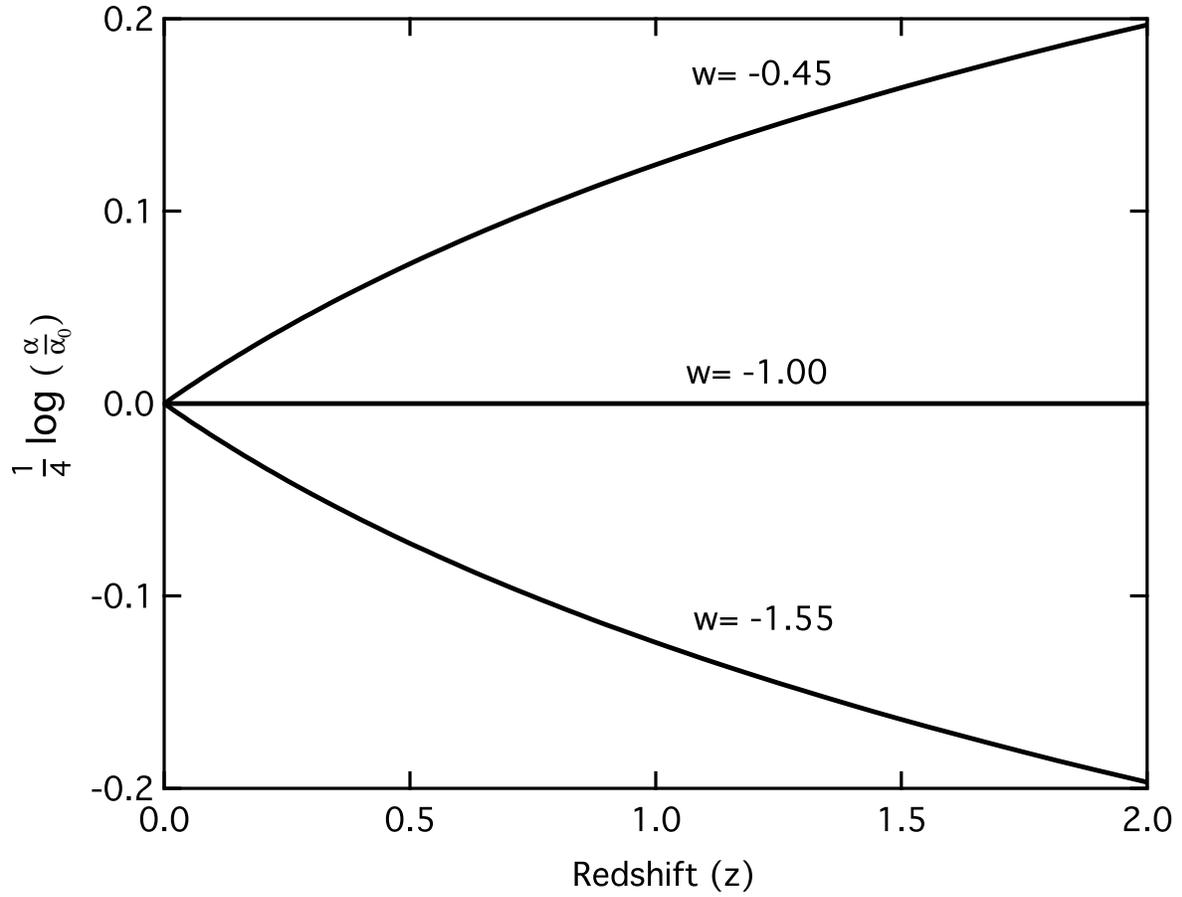}
\caption{Redshift evolution of the MOND acceleration scale, normalized to the
present value $\alpha_0$, if it evolves coupled to
the density of dark energy for different values of the dark 
energy equation of state $w$.}
\label{fig:dark}
\end{figure*}

\begin{figure*}
\centering
\includegraphics[angle=0.0,scale=0.6]{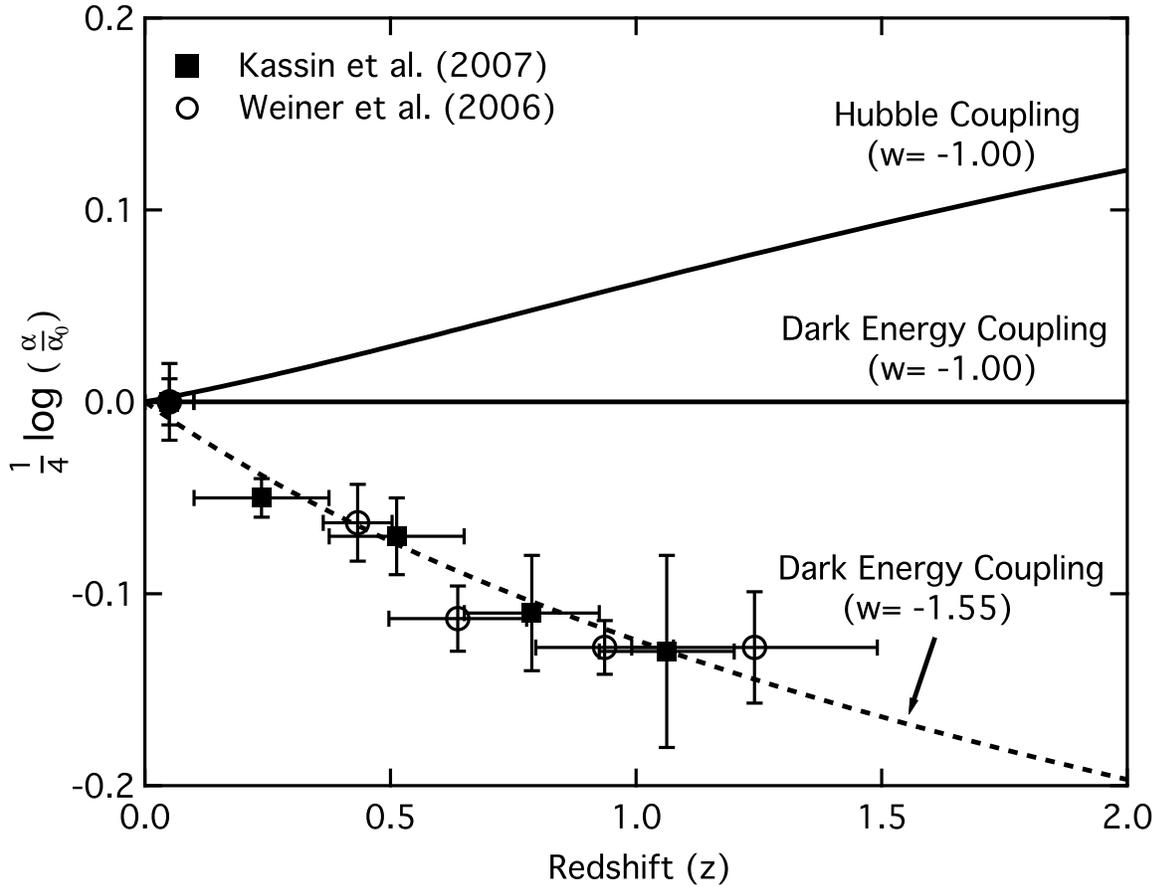}
\caption{The redshift evolution of the MOND acceleration scale, normalized
to the present value $\alpha_0$, if it evolves coupled to the Hubble
parameter or the density of dark energy for different values of the
equation-of-state parameter $w$. The values of the MOND
acceleration scale inferred from infrared J band Tully-Fisher
relations by Weiner et al. (2006) are shown as open circles, where us
those inferred from stellar mass TFR data from Kassin et al. (2007)
are shown as filled squares. We adopt local Tully-Fisher intercept
values from Watanabe (2001) and McGaugh (2005), respectively.}
\label{all}
\end{figure*}


\begin{thebibliography}

%\bibitem[Aguirre(2004)]{2004IAUS..220...17A} Aguirre, A.\ 2004, Dark Matter in Galaxies, 220, 17 

\bibitem[Bekenstein 
\& Sagi(2008)]{2008arXiv0802.1526B} Bekenstein, J.~D., \& Sagi, E.\ 2008, \prd, 77, 103512

\bibitem[Bekenstein \& Milgrom(1984)]{1984ApJ...286....7B} Bekenstein, J., \& Milgrom, M.\ 1984, \apj, 286, 7 

\bibitem[Clowe et al.(2006)]{2006ApJ...648L.109C} Clowe, D., Brada{\v c}, 
M., Gonzalez, A.~H., Markevitch, M., Randall, S.~W., Jones, C., 
\& Zaritsky, D.\ 2006, \apjl, 648, L109

\bibitem[Diemand et al.(2007)]{2007ApJ...667..859D} Diemand, J., Kuhlen, 
M., \& Madau, P.\ 2007, \apj, 667, 859 

\bibitem[Kassin et al.(2007)]{2007ApJ...660L..35K} Kassin, S.~A., et al.\ 
2007, \apjl, 660, L35 

\bibitem[Kaplinghat \& Turner(2002)]{2002ApJ...569L..19K} Kaplinghat, M., \& Turner, M.\ 2002, \apjl, 569, L19 

%\bibitem[Mao et al.(1998)]{1998MNRAS.297L..71M} Mao, S., Mo, H.~J., 
%\& White, S.~D.~M.\ 1998, \mnras, 297, L71 

%\bibitem[McGaugh et al.(2000)]{2000ApJ...533L..99M} McGaugh, S.~S., 
%Schombert, J.~M., Bothun, G.~D., \& de Blok, W.~J.~G.\ 2000, \apjl, 533, L99

\bibitem[McGaugh(2005)]{2005ApJ...632..859M} McGaugh, S.~S.\ 2005, \apj, 
632, 859 

\bibitem[Milgrom(1984)]{1984ApJ...287..571M} Milgrom, M.\ 1984, \apj, 287, 
571 

\bibitem[Milgrom(2001)]{2001AcPPB..32.3613M} ---------.\ 2001, Acta 
Physica Polonica B, 32, 3613

\bibitem[Milgrom(2002)]{2002ApJ...571L..81M} ---------.\ 2002a, \apjl, 571, 
L81 

\bibitem[Milgrom(2002)]{2002NewAR..46..741M} ---------.\ 2002b, New 
Astronomy Review, 46, 741 

\bibitem[Milgrom(2006)]{2006EAS....20..217M} ---------.\ 2006, EAS 
Publications Series, 20, 217, also astro-ph/0510117

\bibitem[Milgrom(2008)]{2008arXiv0801.3133M} ---------.\ 2008, ArXiv 
e-prints, 801, arXiv:0801.3133 

%\bibitem[Navarro et al.(1997)]{1997ApJ...490..493N} Navarro, J.~F., Frenk, 
%C.~S., \& White, S.~D.~M.\ 1997, \apj, 490, 493 

\bibitem[Peebles (1993)]{Peeble 1993} Peebles, P.\,J.\,E. 1993,
Principles of Physical Cosmology

\bibitem[Portinari \& Sommer-Larsen(2007)]{2007MNRAS.375..913P} Portinari, L., \& Sommer-Larsen, J.\ 2007, \mnras, 375, 913

\bibitem[Rubin et al.(1985)]{1985ApJ...289...81R} Rubin, V.~C., Burstein, 
D., Ford, W.~K., Jr., \& Thonnard, N.\ 1985, \apj, 289, 81 

\bibitem[Sanders \& McGaugh (2002)]{2002ARA&A..40..263S} Sanders,
R.~H., \& McGaugh, S.~S.\ 2002, \araa, 40, 263

\bibitem[Setal01]{Setal01} Scott, D., White, M., Cohn, J.\,D., \& Pierpaoli,
E.\, astro-ph/0104435

\bibitem[Spergel et al.(2007)]{2007ApJS..170..377S} Spergel, D.~N., et al.\ 
2007, \apjs, 170, 377 

\bibitem[Tully \& Fisher(1977)]{1977A&A....54..661T} Tully, R.~B., \& Fisher, J.~R.\ 1977, \aap, 54, 661 

\bibitem[Verheijen(2001)]{2001ApJ...563..694V} Verheijen, M.~A.~W.\ 2001, 
\apj, 563, 694 

\bibitem[Watanabe et al.(2001)]{2001ApJ...555..215W} Watanabe, M., Yasuda, 
N., Itoh, N., Ichikawa, T., \& Yanagisawa, K.\ 2001, \apj, 555, 215 

\bibitem[Weiner et al.(2006)]{2006ApJ...653.1049W} Weiner, B.~J., et al.\ 
2006, \apj, 653, 1049 

\bibitem[Zhao \& Li(2008)]{2008arXiv0804.1588Z} Zhao, H., \& Li, B.\ 2008, ArXiv e-prints, 804, arXiv:0804.1588 

\end{thebibliography}
\end{document}